\documentclass[runningheads]{llncs} 

\usepackage[T1]{fontenc}

\let\oldfootnote=\footnote
\def\footnote{\suppressfloats[b]\oldfootnote}

\usepackage[hidelinks]{hyperref}
\usepackage{xurl}
\usepackage{tablefootnote}
\usepackage{comment}
\usepackage{flushend}
\usepackage{amsmath} 

\usepackage{academicons}
\usepackage{xcolor}
\newcommand{\orcid}[1]{\href{https://orcid.org/#1}{\textcolor[HTML]{A6CE39}{\aiOrcid}}}

\usepackage[font=scriptsize,labelfont=scriptsize]{caption}
\usepackage{booktabs} 
\usepackage{graphicx} 
\usepackage{enumitem}
\usepackage{subcaption} 
\setlist[itemize]{noitemsep, topsep=0pt}

\newcommand{\PP}{\reflectbox{P}\hspace{-.1cm}\text{P}}

\let\originalbottomrule\bottomrule
\renewcommand{\bottomrule}{\addlinespace[0pt]\originalbottomrule}
\let\originalmidrule\midrule
\renewcommand{\midrule}{\addlinespace[0pt]\originalmidrule}

\begin{document}

\title{Bringing Auto-tuning to HIP: Analysis of Tuning Impact and Difficulty on AMD and Nvidia GPUs}
\titlerunning{Bringing auto-tuning to HIP}

\author{Milo Lurati\inst{1,2} \and
Stijn Heldens\inst{2}\orcidID{0000-0001-8792-6305} \and
Alessio Sclocco\inst{2}\orcidID{0000-0003-3278-0518} \and
Ben van Werkhoven\inst{3,2}\orcidID{0000-0002-7508-3272}}
\authorrunning{M. Lurati et al.}

\institute{VU Amsterdam, Netherlands \and
Netherlands eScience Center, Amsterdam, Netherlands \and
Leiden Institute of Advanced Computer Science, Leiden, Netherlands
}
\maketitle              
%
%

\begin{abstract}
Many studies have focused on developing and improving auto-tuning algorithms for Nvidia Graphics Processing Units (GPUs), but the effectiveness and efficiency of these approaches on AMD devices have hardly been studied. This paper aims to address this gap by introducing an auto-tuner for AMD's HIP. We do so by extending Kernel Tuner, an open-source Python library for auto-tuning GPU programs. We analyze the performance impact and tuning difficulty for four highly-tunable benchmark kernels on four different GPUs: two from Nvidia and two from AMD. Our results demonstrate that auto-tuning has a significantly higher impact on performance on AMD compared to Nvidia (10x vs 2x). Additionally, we show that applications tuned for Nvidia do not perform optimally on AMD, underscoring the importance of auto-tuning specifically for AMD to achieve high performance on these GPUs.
\end{abstract}

\setlength{\abovedisplayskip}{0pt}
\setlength{\belowdisplayskip}{0pt}

\section{Introduction}

Graphics Processing Units (GPUs) are widely used in High-Performance Computing (HPC) and artificial intelligence because of their high parallel processing power and ability to accelerate complex workloads~\cite{heldens2020landscape,lecun2015deep}. Eight out of nine supercomputers funded by EuroHPC JU use GPUs as the main source of compute power\footnote{\url{https://eurohpc-ju.europa.eu/supercomputers/our-supercomputers_en} (Accessed March 2024)}. GPUs excel in terms of compute performance and energy efficiency for tasks that involve large data sets and dense computation, making them increasingly vital in various scientific domains~\cite{van2020lessons}. 

GPU programming models -- such as HIP, CUDA, and OpenCL -- allow developers to create highly parallel functions, called {\em kernels}. However, GPU programmers are confronted with a myriad of implementation choices and optimization techniques related to thread organization, memory usage, and computation strategies to achieve optimal compute performance~\cite{hijma2023optimization}. The optimal kernel configuration depends on the specific GPU architecture and the task at hand, and finding this configuration is a process known as performance tuning; automating this process is called auto-tuning~\cite{balaprakash2018autotuning}.

While auto-tuning techniques have been extensively studied for Nvidia GPUs \cite{li2009note,agullo2009numerical,nath2010improved,magni2013input,rasch2021efficient,nukada2009auto,grauer2012auto,van2014optimizing}, their effectiveness on AMD GPUs has received considerably less attention. The studies that do consider AMD GPUs predominantly use OpenCL~\cite{sclocco2014,sclocco2020amber,van2019kernel}. 
In 2016, AMD introduced HIP: an open-source GPU programming model that enables applications to run on both AMD and Nvidia GPUs through a single source code. 
HIP creates new opportunities for auto-tuning. For example, OpenCL on AMD was restricted to at most 256 threads per block~\cite{komatsu2010evaluating,dolbeau2013one,yu2019efficient,schoonhoven2022benchmarking}, whereas HIP increases this limit to 1024. 

After a long period of market dominance by Nvidia, the HPC landscape is rapidly diversifying with the first generation of exascale supercomputers featuring for example Intel~\cite{aurora} and AMD GPUs~\cite{frontier}. Europe's \#1 supercomputer LUMI, which uses AMD's MI250X GPUs, is part of this trend. It is urgent that we understand how the lessons learned from optimizing and tuning applications predominantly on Nvidia GPUs for over a decade, can be migrated to GPUs from different vendors. 

To this end, this paper introduces the first auto-tuning tool for HIP kernels and studies the performance impact of tuning HIP kernels on AMD GPUs. 
Since HIP applications can run on both AMD and Nvidia GPUs, we subsequently compare the impact, tuning difficulty, and performance portability of tuned HIP applications on both AMD and Nvidia GPUs.

\noindent The contributions of this work are as follows:\begin{itemize}[noitemsep,topsep=-2pt]
    \item We extend Kernel Tuner~\cite{van2019kernel}, an open-source Python tool for auto-tuning GPU applications, with support for HIP by integrating PyHIP, an open-source Python library to access the HIP runtime library and compiler~\cite{PyHIP}.
    \item We compare performance and portability of four highly-optimized auto-tuned HIP kernels on two AMD and two Nvidia GPUs.
    \item We show that GPUs by Nvidia are generally easier to tune than those from AMD, both manually and using optimization algorithms, while the performance impact of tuning the same code on AMD GPUs is much larger compared to Nvidia (10x vs 2x).
    \item We show that kernels tuned for AMD generally perform well on Nvidia GPUs, but not the other way around. 
\end{itemize}

\noindent These findings demonstrate that it is even more important to use auto-tuning for HIP applications on AMD GPUs, compared to Nvidia, and thus emphasize the need for new tools that enable auto-tuning HIP code for AMD GPUs.

\section{Related Work}\label{sec:related_work}

Auto-tuning is widely used in various contexts such as optimizing numerical libraries, compilers, and application performance~\cite{balaprakash2018autotuning}. Examples of applications using auto-tuning include FFTW~\cite{frigo1998fftw} for optimizing Fast Fourier Transforms on CPUs~\cite{vuduc2000code} and MAGMA for linear algebra~\cite{agullo2009numerical}.
In this paper, we focus on software-level auto-tuning, and in particular on the automatic tuning of code that targets GPUs.

There are several generic auto-tuners targeting GPU code.
CLTune~\cite{nugteren2015cltune} is an auto-tuner for OpenCL. KTT~\cite{filipovivc2017autotuning} tunes parameters in OpenCL, CUDA, and GLSL applications focusing on pipelines of multiple kernels. ATF~\cite{rasch2021efficient} focuses on OpenCL and CUDA kernels that have interdependent parameters. 

HIP was released by AMD in March 2016 and is increasingly being adopted as a programming model for HPC applications, such as AMBER\footnote{\url{https://ambermd.org/GPUSupport.php}}, NAMD\footnote{\url{http://www.ks.uiuc.edu/Research/namd/alpha/2.15_amdhip/}}, PeleC\footnote{\url{https://amrex-combustion.github.io/PeleC/}}, and AMReX\footnote{\url{https://amrex-codes.github.io/amrex/docs_html/GPU.html}}.
However, HIP is, to the best of our knowledge, not supported by any current auto-tuning framework. 

In general, most auto-tuning studies have focused primarily on auto-tuning applications for Nvidia GPUs~\cite{li2009note,nath2010improved,magni2013input,rasch2021efficient,nukada2009auto,grauer2012auto,van2014optimizing,filipovivc2017autotuning,torring2023towards}. 
Many auto-tuning studies have included one or more AMD GPUs using OpenCL~\cite{komatsu2010evaluating,nugteren2015cltune,hou2017auto,yu2019efficient,van2019kernel,schoonhoven2022benchmarking}. 
To the best of our knowledge, this paper is the first study to investigate and compare the impact, tuning difficulty, and performance portability on both AMD and Nvidia GPUs for auto-tuned HIP applications.

\section{Design and Implementation}\label{sec:design}

\begin{figure}[t!]
  \centering
  \vspace{-0.5cm}
  \begin{minipage}[t]{0.5\textwidth}
  \includegraphics[width=\linewidth]{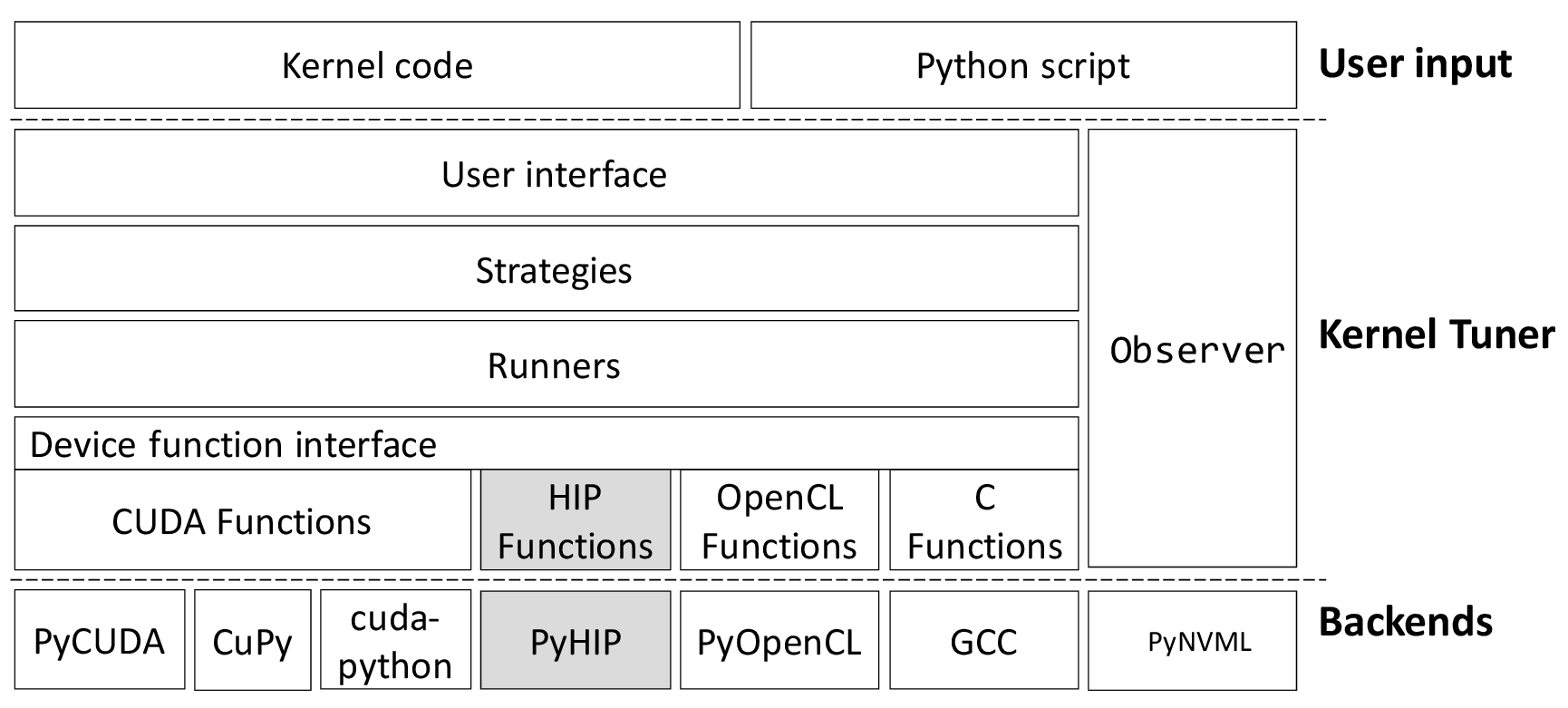}
  \vspace{-0.7cm}
  \caption{Kernel Tuner software architecture.}
  \label{fig:Architecture}
  \end{minipage}%
  \begin{minipage}[t]{0.499\textwidth}
    \centering
    \includegraphics[width=0.9\linewidth]{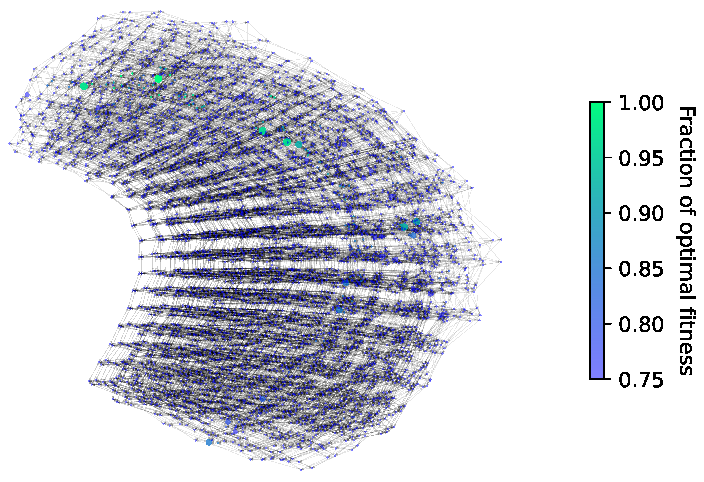}
    \vspace{-0.4cm}
    \caption{Fitness Flow Graph of 2D Convolution search space for A4000.}
    \label{fig:FFG_A4000_conv}
  \end{minipage}
  \vspace{-0.5cm}
\end{figure}

The layered software architecture of Kernel Tuner, extended to accommodate our contributions, is shown in Figure~\ref{fig:Architecture}. This revised architecture incorporates the HIP functions interface built on top of PyHIP\footnote{\url{https://github.com/jatinx/PyHIP}}. 

Users of Kernel Tuner create a small Python script that describes how the GPU code can be tuned. 
The strategies layer implements a great variety of optimization algorithms, which in turn rely on a runner. The runners interact with the diverse set of supported compilers and hardware through a unified device function interface, which abstracts the device-specific functionalities offered by various backends such as PyCUDA, CuPy, and PyHIP. This allows the higher-level layers (e.g. runners, optimization strategies) to operate independently of the underlying hardware and runtime.

The HIP backend in Kernel Tuner builds on PyHIP, a Python wrapper for HIP. We have made various contributions to PyHIP to increase its coverage of the HIP Runtime API and simplified the installation procedure.
To integrate the new HIP backend with the rest of Kernel Tuner several changes were made. Due to the very high similarity between CUDA and HIP kernels, Kernel Tuner is not able to automatically detect the kernel language. To solve this problem, we require the user to manually specify when HIP is used.

Kernel Tuner performs empirical measurements of the execution time of each kernel configuration it compiles and benchmarks. As with CUDA, the execution time of HIP kernels is measured by recording events before and after the kernel and calling {\tt hipEventElapsedTime} to retrieve the execution time. 

Finally, to support loop-unrolling, a code optimization that aims to improve program performance by reducing loop overhead, while increasing instruction-level parallelism~\cite{hijma2023optimization}, we have extended support in Kernel Tuner to auto-tuning partial loop unrolling factors in CUDA kernels to also support HIP kernels.

\section{Evaluation metrics}
\label{sec:tuning}

We compare auto-tuning GPU codes for either vendor along three main axes: performance impact of auto-tuning, the tuning difficulty, and the performance portability of tuned kernels.

\noindent \textbf{Tuning impact}.
To quantify the performance impact of auto-tuning we analyze the statistical properties of the performance distribution of the full tuning search space of a kernel.
More specifically, we define {\em tuning impact} as the factor between the performance of the {\em global optimum} and the {\em median} performance of configurations in the space. The rationale is that without auto-tuning one can expect to achieve performance that is the most common among configurations, and with auto-tuning the application can achieve optimal performance.
In addition, violin plots are used to visualize the performance distributions relative to the optimum across devices, allowing for direct comparison and pattern identification.

\noindent \textbf{Tuning difficulty}. For some tuning spaces, the global optimum may be a statistical outlier in terms of performance, but that does not necessarily mean that the global optimum is also difficult to find for an optimization algorithm. To assess the respective tuning difficulty on GPUs from the different vendors, we quantify how difficult it is for an optimization algorithm to arrive at a configuration of acceptable performance. For this, we use \emph{the proportion of centrality}~\cite{schoonhoven2022benchmarking}. 

The proportion of centrality is computed on a fitness flow graph (FFG),
which has directed edges
between neighbouring points with better fitness values, as shown in Figure~\ref{fig:FFG_A4000_conv}. The idea is that a random walk on the FFG simulates the path taken by a local search algorithm. We use PageRank~\cite{brin1998anatomy} centrality to quantify the likelihood of arriving at a local minimum.
Given a proportion \(p\), consider \(f_{opt}\) as the optimal fitness, \(L(X)\) as the set of local minima of \(X\), and \(L_{p}(X)\) as the collection of local minima with fitness values less than \((1 + p)f_{opt}\). P-proportion of centrality is defined, with \(c_{G}\) as the centrality function, as:\begin{equation}
    C_{p}(G,X) = \frac{\sum_{x\in L_{p}(X)}c_{G}(x)}{\sum_{x\in L(X)}c_{G}(x)} \label{eq:prop_centr}
\end{equation}

\noindent \textbf{Performance portability}.
Performance portability examines how well a configuration that gives optimal performance on one device or set of devices, performs when moving to another device.
We use the metric defined by Pennycook et al.~\cite{pennycook2016metric}, denoted as \PP{}, which measures the performance portability across a set of devices $H$ for configuration $x$ of kernel $p$ as:\newline
\begin{minipage}[]{0.499\textwidth}
    \begin{equation}
    \PP(x, p, H) = \frac{|H|}{\sum_{i \in H} \frac{1}{e_i(x, p)}}\label{eq:performance_portability}
    \end{equation}
\end{minipage}%
\begin{minipage}[]{0.499\textwidth}
    \begin{equation}
    e_i(x, p) = \frac{P_i(x, p)}{ \underset{x \in X}{\max}~P_i(x, p) } \label{eq:application_efficiency}
    \end{equation}
\end{minipage}

\noindent Here, $e_i(x, p)$ represents the \emph{performance efficiency} of configuration $x$ for kernel $p$ on device $i$ as the ratio of the achieved performance $P_i(x, p)$ to the highest observed performance across all configurations called $X$.

\section{Experimental setup}\label{sec:setup}

In this section, we introduce the benchmark applications and the hardware and software used to compare auto-tuning HIP code on AMD and Nvidia GPUs.

\noindent \textbf{Benchmark kernels}.
For the evaluation, we use four benchmark kernels taken from the CLBlast library~\cite{nugteren2018clblast} (namely GEMM) and the BAT benchmark suite~\cite{torring2023towards} (namely Convolution, Hotspot, and Dedispersion). The problems implemented by these kernels and an explanation of their tunable parameters can be found in \cite{nugteren2018clblast} and \cite{torring2023towards}. The tunable parameter values are listed in Table~\ref{tab:parameters_conv_hot_dedisp} and \ref{tab:parameters_gemm}. For GEMM, the input matrices are 4096x4096. The full source code of the kernels, input problem dimensions, and analysis tools are provided in the accompanying GitHub repository\footnote{\url{https://github.com/MiloLurati/AutoTuning\_AMD\_vs\_Nvidia\_GPUs}}.

\noindent \textbf{Hardware and software description}.
For the evaluation, we focus on four different GPU models available in the DAS-6 cluster and the LUMI supercomputer.
The GPU specifications are listed in Table~\ref{tab:gpu-properties}. On DAS-6 we use Rocky-8 Linux 4.18.0, ROCM 6.0.2 with AMD clang 17.0.0, and CUDA 12.2 with GCC 9.4.0. For the MI250X, LUMI is running SUSE Linux 5.14.21, ROCM 5.2.3 with AMD clang 14.0.0. 
Note that the MI250X is a multi-chip module with two individually operating GPU dies and we use only a single die.
All measurements have been performed with Kernel Tuner 1.0.0b6, into which our modifications have been merged.
For proportion of centrality calculation and visualization, we adapted the code from Schoonhoven et al.~\cite{schoonhoven2022benchmarking}.

\begin{table}[tb]
    \centering
    \tiny
    \vspace{-0.5cm}
    \caption{Tunable parameters for Convolution, Hotspot, and Dedispersion kernels.}\label{tab:parameters_conv_hot_dedisp}
    \begin{tabular}{l|l|l|l}
    \toprule
Parameter	& Convolution	& Hotspot	& Dedispersion	\\
\midrule
block\_size\_x	& $16k$ for $k$ in $1, 2, \ldots, 16$	& $1, 2, 4, 8, 16, \newline 32k$ for $k$ in $1, 2, \ldots, 32$	& {1, 2, 4, 8, 16, 32}	\\
block\_size\_y	& {1, 2, 4, 8, 16}	& {1, 2, 4, 8, 16, 32} 	& $8k$ for $k$ in $4, 5, \ldots, 32$	\\
tile\_size\_x	& {1, 2, 3, 4}	& {1, 2, 3, 4, 5, 6, 7, 8, 9, 10} 	& {1, 2, 3, 4}	\\
tile\_size\_y	& {1, 2, 3, 4}	& {1, 2, 3, 4, 5, 6, 7, 8, 9, 10} 	& {1, 2, 3, 4, 5, 6, 7, 8}	\\
read\_only	& {0, 1}	& 	& 	\\
use\_padding	& {0, 1}	& 	& 	\\
use\_shmem	& {0, 1}	& 	& 	\\
temporal\_tiling\_factor	& 	& {1, 2, 3, 4, 5, 6, 7, 8, 9, 10} 	& 	\\
loop\_unroll\_factor\_t	& 	& {1, 2, 3, 4, 5, 6, 7, 8, 9, 10} 	& 	\\
sh\_power	& 	& {0, 1}	& 	\\
tile\_stride\_x 	& 	& 	& {0, 1}	\\
tile\_stride\_y	& 	& 	& {0, 1}	\\
\bottomrule
\end{tabular}
    \vspace{-0.5cm}
\end{table}

\begin{table}[tb]
    \tiny
    \centering
    \caption{GPUs used in our experiments. *Only one out of two dies of MI250X is used.}\label{tab:gpu-properties}
    \begin{tabular}{l|l|l|r|r|r|r|r}
    \toprule
    GPU & Year & Architecture & Cores & Memory & Cache & Bandwidth (GB/s) & Peak SP (GFLOPS/s)\\
    \midrule
    AMD W6600    & 2021 & RDNA 2 & 1792 & 16~GB GDDR6 & 32 MB L3 & 224  & 10404 \\
    AMD MI250X*  & 2021 & CDNA 2 & 7040 & 64~GB HMB2e & 8 MB L2  & 1638 & 28160 \\
    Nvidia A4000 & 2021 & Ampere & 6144 & 8~GB GDDR6  & 4 MB L2  & 448  & 17800 \\
    Nvidia A100  & 2020 & Ampere & 6912 & 40~GB HMB2  & 40 MB L2 & 1555 & 19500 \\
    \bottomrule
    \end{tabular}
    \vspace{-0.5cm}
\end{table}

\section{Evaluation}
\label{sec:evaluation}

In this section, we first present the results on the four benchmark applications by analyzing the tuning impact, tuning difficulty, and performance portability. We also present the top five best performing configurations in each auto-tuning search space to discuss how the results obtained by the tuner can be explained by properties of the hardware.

\subsection{Convolution}

\begin{table}[tb]
\begin{minipage}[t]{0.3\textwidth}
    \centering
    \tiny
    \caption{GEMM tunable parameters, as explained in~\cite{nugteren2018clblast}.}\label{tab:parameters_gemm}
    \begin{tabular}{l|l}
    \toprule
    Parameter & Values \\
    \midrule
    MWG	& {16, 32, 64, 128}	\\
    NWG	& {16, 32, 64, 128}	\\
    KWG	& {16, 32}	\\
    MDIMC	& {8, 16, 32}	\\
    NDIMC	& {8, 16, 32}	\\
    MDIMA	& {8, 16, 32}	\\
    NDIMB	& {8, 16, 32}	\\
    VWM	& {1, 2, 4, 8}	\\
    VWN	& {1, 2, 4, 8}	\\
    STRM	& {0, 1}	\\
    STRN	& {0, 1}	\\
    SA	& {0, 1}	\\
    SB	& {0, 1}	\\
    \bottomrule
    \end{tabular}
\end{minipage}\hspace{0.5cm}%
\begin{minipage}[t]{0.6\textwidth}
    \tiny
    \centering
    \caption{Statistical properties of the benchmarks. \emph{Tuning impact} is the maximum over the median.}
    \label{tab:result_stats}
    \begin{tabular}{l|lrrrr} 
         \toprule
         && W6600 & MI250X & A4000 & A100 \\ 
         \hline
         Convolution (GFLOP/s)
         &median & 137    & 380     & 2284  & 4117 \\
         4,362 configurations
         &maximum& 4370   & 11460   & 7393  & 13637 \\ 
         &impact & 31.9x & 30.1x & 3.2x & 3.3x \\
         \midrule
         Hotspot (GFLOP/s)
         &median & 94     & 334    & 92    & 632 \\ 
         10,5412 configurations
         &maximum& 229    & 1781   & 177   & 1776 \\ 
         &impact& 2.5x   & 5.3x   & 1.9x  & 2.8x \\
         \midrule
         Dedispersion (GB/s)
         &median & 427    & 667     & 470     & 1085 \\ 
         11,130 configurations
         &maximum& 582    & 1586    & 532     & 1154 \\ 
         &impact& 1.4x   & 2.4x    & 1.1x    & 1.1x \\
         \midrule
         GEMM (GFLOPS/s)
         &median & 1154   & 7799    & 4802    & 10748 \\ 
         116,928 configurations
         &maximum& 6010  & 19807   & 10502   & 17145 \\ 
         &impact& 5.2x   & 2.5x    & 2.2x    & 1.6x \\
         \bottomrule
    \end{tabular}
\end{minipage}
\vspace{-0.5cm}
\end{table}

\begin{figure}[tb]
\vspace{-0.2cm}
    \centering
    \begin{minipage}[t]{0.499\textwidth}
        \includegraphics[width=\linewidth]{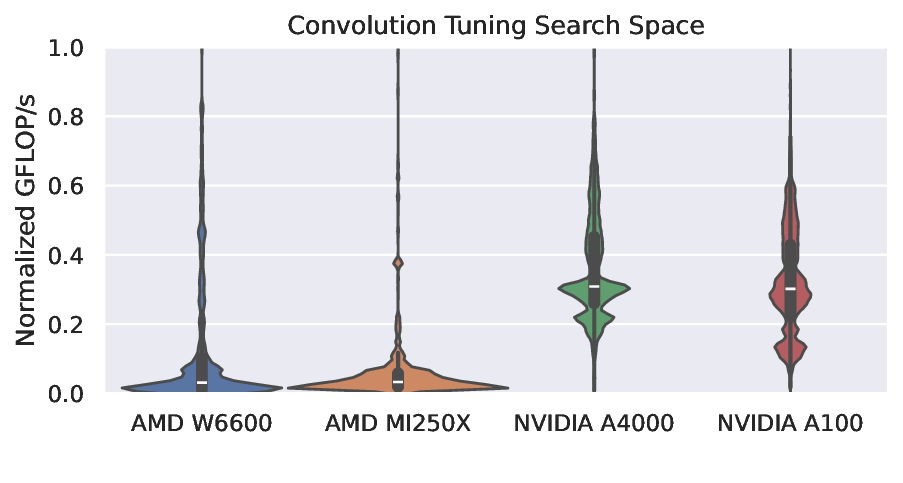}
        \vspace{-0.7cm}
        \caption{2D Convolution tuning search space.}
        \label{fig:Conv_SP_violins}
    \end{minipage}%
    \begin{minipage}[t]{0.499\textwidth}    
        \includegraphics[width=\linewidth]{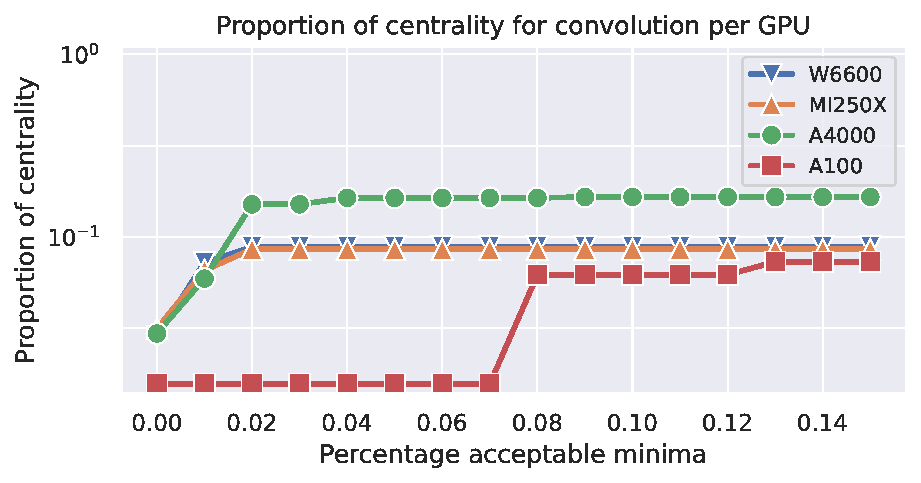}
        \vspace{-0.7cm}
        \caption{2D Convolution proportion of centrality.}
        \label{fig:prop_centr_conv}
    \end{minipage}    
\vspace{-0.2cm}
    \scriptsize
\captionof{table}{Top configurations for convolution. Parameters match Table~\ref{tab:parameters_conv_hot_dedisp}. Performance in TFLOP/s.}\label{conv_params}
\small
\centering
\begin{minipage}{0.249\columnwidth}\scriptsize\centering
W6600 \\
\begin{tabular}{rrrrrrr|r}
\toprule
\multicolumn{7}{c|}{Parameters} & Perf. \\
\midrule
128 & 1 & 1 & 4 & 1 & 0 & 0 & 4.37 \\ 
32 & 1 & 1 & 4 & 1 & 0 & 0 & 4.35 \\ 
64 & 1 & 1 & 4 & 1 & 0 & 0 & 4.33 \\ 
256 & 1 & 1 & 4 & 1 & 0 & 0 & 4.32 \\ 
16 & 16 & 4 & 2 & 1 & 1 & 1 & 3.65 \\ 
\midrule
\end{tabular} \\
\end{minipage}%
\begin{minipage}{0.249\columnwidth}\scriptsize\centering
MI250X \\
\begin{tabular}{rrrrrrr|r}
\toprule
\multicolumn{7}{c|}{Parameters} & Perf. \\
\midrule
64 & 1 & 2 & 4 & 1 & 0 & 0 & 11.46 \\ 
128 & 1 & 2 & 4 & 1 & 0 & 0 & 11.46 \\ 
256 & 1 & 2 & 4 & 1 & 0 & 0 & 11.29 \\ 
128 & 1 & 1 & 4 & 1 & 0 & 0 & 11.28 \\ 
64 & 1 & 1 & 4 & 1 & 0 & 0 & 11.23 \\ 
\midrule
\end{tabular} \\
\end{minipage}%
\begin{minipage}{0.249\columnwidth}\scriptsize\centering
A4000 \\
\begin{tabular}{rrrrrrr|r}
\toprule
\multicolumn{7}{c|}{Parameters} & Perf. \\
\midrule
256 & 1 & 2 & 4 & 0 & 0 & 0 & 7.39 \\ 
32 & 1 & 2 & 4 & 0 & 0 & 0 & 7.36 \\ 
128 & 1 & 2 & 4 & 0 & 0 & 0 & 7.31 \\ 
256 & 1 & 1 & 4 & 0 & 0 & 0 & 7.30 \\ 
32 & 1 & 4 & 4 & 0 & 0 & 0 & 7.30 \\ 
\midrule
\end{tabular} \\
\end{minipage}%
\begin{minipage}{0.249\columnwidth}\scriptsize\centering
A100 \\
\begin{tabular}{rrrrrrr|r}
\toprule
\multicolumn{7}{c|}{Parameters} & Perf. \\
\midrule
32 & 4 & 1 & 3 & 1 & 0 & 1 & 13.64 \\ 
128 & 2 & 1 & 3 & 1 & 0 & 1 & 12.69 \\ 
128 & 1 & 1 & 3 & 1 & 0 & 1 & 12.27 \\ 
48 & 2 & 1 & 4 & 1 & 0 & 1 & 12.09 \\ 
48 & 2 & 1 & 3 & 1 & 0 & 1 & 12.08 \\ 
\midrule
\end{tabular} \\
\end{minipage}%
\end{figure}

Figure~\ref{fig:Conv_SP_violins} presents the performance distributions of the convolution kernel tuning space on all four GPUs showing rather bottom-heavy distributions, meaning that the optimal configurations are extreme outliers in terms of performance. 
This is, however, even more pronounced for the two AMD devices.
It is quite clear from these results that manual performance optimization of the convolution kernel is, if not impossible, at least very unlikely to result in optimal performance.

The median and maximum of each kernel on each device are shown in Table~\ref{tab:result_stats}, showing how important tuning is for this kernel, in absolute performance: tuning provides a ${\sim}$30x performance improvement for the AMD GPUs, and a ${\sim}$3x improvement for the Nvidia ones.
A whole order of magnitude difference between the two vendors, meaning the impact of auto-tuning is high for our AMD devices.

Figure~\ref{fig:prop_centr_conv} shows the proportion of centrality of the convolution, for all platforms, at different levels of acceptable optima $p$, ranging from 0\% (the global optimum) to 15\%.
Here we see that, while manual tuning was more difficult for the AMD GPUs, the results for this experiment are different.
Instead of a vendor split, we see that finding the global optimum of the A100 is more difficult than finding the optimum of the other devices, and that by relaxing the constraints on the optimum the A4000 becomes easier to tune than the rest.

Table~\ref{conv_params} shows the top 5 configurations for each device. A first observation is that these configurations are different for each device.
However, we can observe certain patterns.
All GPUs prefer small thread blocks, with at most 256 threads, but while the two AMD devices, and the A4000, prefer one-dimensional block configurations, the A100 prefers two-dimensional ones.
So, even if the total number of threads is similar, the distribution of threads in the two-dimensional block is not.
Another similarity between the GPUs is that all configurations use some form of tiling in the $y$ dimension, to compensate for the lack 
of thread-level parallelism within thread blocks.
In contrast, tiling in the $x$ dimension is mainly used by the MI250X and the A4000, and not by the other two devices.
Two more facts to highlight are that the A100 is the only GPU to consistently prefer using shared memory, but without padding to avoid bank conflicts, which is only used by one configuration in the top 5 on the W6600 with a 16x16 thread block size.

\subsection{Hotspot}

\begin{figure}[tb]
\vspace{-0.2cm}
    \centering
    \begin{minipage}[t]{0.499\textwidth}
        \includegraphics[width=\linewidth]{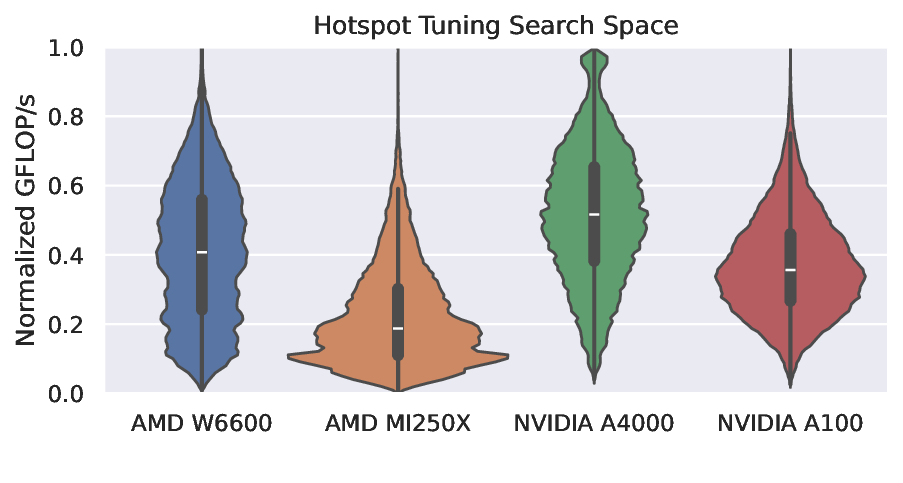}
        \vspace{-0.7cm}
        \caption{Hotspot tuning search space.}
        \label{fig:HS_SP_violins}
    \end{minipage}%
    \begin{minipage}[t]{0.499\textwidth}
        \includegraphics[width=\linewidth]{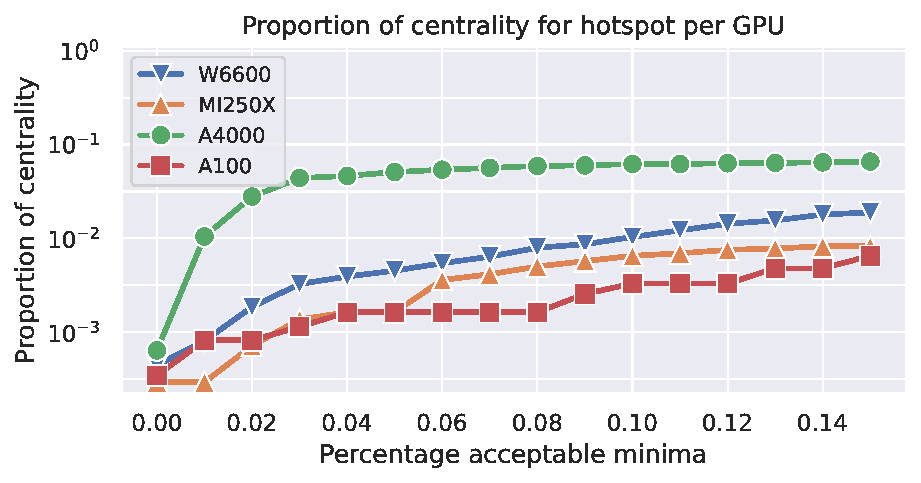}
        \vspace{-0.7cm}
        \caption{Hotspot proportion of centrality.}
        \label{fig:prop_centr_hotspot}
    \end{minipage}
\vspace{-0.2cm}
    \scriptsize
\captionof{table}{Top configurations for Hotspot. Parameters match Table~\ref{tab:parameters_conv_hot_dedisp}. Performance in GFLOP/s.}\label{hotspot_params}
\small
\centering
\begin{minipage}{0.249\columnwidth}
\scriptsize
\centering
W6600 \\
\begin{tabular}{rrrrrrr|r}
\toprule
\multicolumn{7}{c|}{Parameters} & Perf. \\
\midrule
8 & 32 & 4 & 2 & 3 & 3 & 1 & 229.31 \\ 
16 & 16 & 4 & 2 & 3 & 3 & 1 & 227.13 \\ 
8 & 32 & 8 & 1 & 3 & 3 & 1 & 226.72 \\ 
8 & 32 & 7 & 1 & 4 & 4 & 1 & 226.70 \\ 
16 & 32 & 4 & 1 & 3 & 3 & 1 & 226.49 \\ 
\midrule
\end{tabular} \\
\end{minipage}%
\begin{minipage}{0.249\columnwidth}\scriptsize\centering
MI250X  \\
\begin{tabular}{rrrrrrr|r}
\toprule
\multicolumn{7}{c|}{Parameters} & Perf. \\
\midrule
16 & 32 & 2 & 1 & 5 & 5 & 1 & 1781.16 \\ 
16 & 32 & 2 & 1 & 4 & 4 & 1 & 1738.68 \\ 
32 & 32 & 4 & 1 & 4 & 4 & 1 & 1723.78 \\ 
32 & 16 & 2 & 1 & 4 & 4 & 1 & 1690.77 \\ 
16 & 32 & 6 & 1 & 8 & 8 & 1 & 1685.52 \\ 
\midrule
\end{tabular} \\
\end{minipage}%
\begin{minipage}{0.249\columnwidth}\scriptsize\centering
\scriptsize
\centering
A4000 \\
\begin{tabular}{rrrrrrr|r}
\toprule
\multicolumn{7}{c|}{Parameters} & Perf. \\
\midrule
64 & 1 & 8 & 2 & 1 & 1 & 0 & 177.93 \\ 
64 & 2 & 8 & 2 & 1 & 1 & 0 & 177.59 \\ 
32 & 2 & 8 & 3 & 1 & 1 & 0 & 177.40 \\ 
64 & 2 & 8 & 3 & 1 & 1 & 0 & 177.31 \\ 
32 & 2 & 4 & 8 & 1 & 1 & 0 & 177.29 \\ 
\midrule
\end{tabular}
\end{minipage}%
\begin{minipage}{0.249\columnwidth}\scriptsize\centering
A100 \\
\begin{tabular}{rrrrrrr|r}
\toprule
\multicolumn{7}{c|}{Parameters} & Perf. \\
\midrule
8 & 32 & 4 & 1 & 4 & 4 & 1 & 1776.14 \\ 
4 & 32 & 4 & 1 & 9 & 1 & 1 & 1770.40 \\ 
4 & 32 & 5 & 2 & 7 & 1 & 1 & 1763.47 \\ 
8 & 32 & 4 & 1 & 4 & 2 & 1 & 1747.24 \\ 
8 & 32 & 6 & 1 & 4 & 4 & 1 & 1741.92 \\ 
\midrule
\end{tabular} \\
\end{minipage}
\end{figure}

Next, we study the Hotspot kernel.
Figure~\ref{fig:HS_SP_violins} shows a clear separation between consumer and server grade GPUs, with the consumer GPUs having more configurations that lead to reasonably good performance, and the server grade GPUs showing that only a few configurations achieve high performance. As shown in Table~\ref{tab:result_stats}, the impact of auto-tuning the hotspot kernel varies from 1.9x on the A4000 to 5.3x on the MI250X.

The consumer grade GPUs are also easier to tune for optimization algorithms, as shown in Figure~\ref{fig:prop_centr_hotspot}, although in this case the tuning difficulty of the two AMD devices is not that different from each other once we relax the amount of acceptable configurations.

Table~\ref{hotspot_params} shows the top 5 configurations on all four devices. Again, we see that no configuration appears twice, underlining the need to tune for each device individually.
The A4000 stands out, it is has the worst performance of all four GPUs and is the only GPU that does not store the power input data in shared memory. Also, the A4000 does not use temporal tiling, and instead uses relatively small block sizes combined with spatial tiling. All to reduce register usage and improve thread-level parallelism at the cost of data reuse in shared memory.

The other GPUs all use some degree of temporal tiling, which computes multiple calls of the kernel in a single kernel call, trading increased SM-level resource usage and even redundant work for reduced DRAM traffic. The AMD GPUs prefer to fully unroll the temporal tiling loop, where this preference is less pronounced on the A100.
The MI250X uses large thread blocks, up to 1024 threads, much larger than the A100, showing that while the distributions, and even performance, of the two devices are similar, the optimal configurations are not.

\subsection{Dedispersion}

\begin{figure}[tb]
\vspace{-0.2cm}
    \centering
    \begin{minipage}[t]{0.499\textwidth}
        \includegraphics[width=\linewidth]{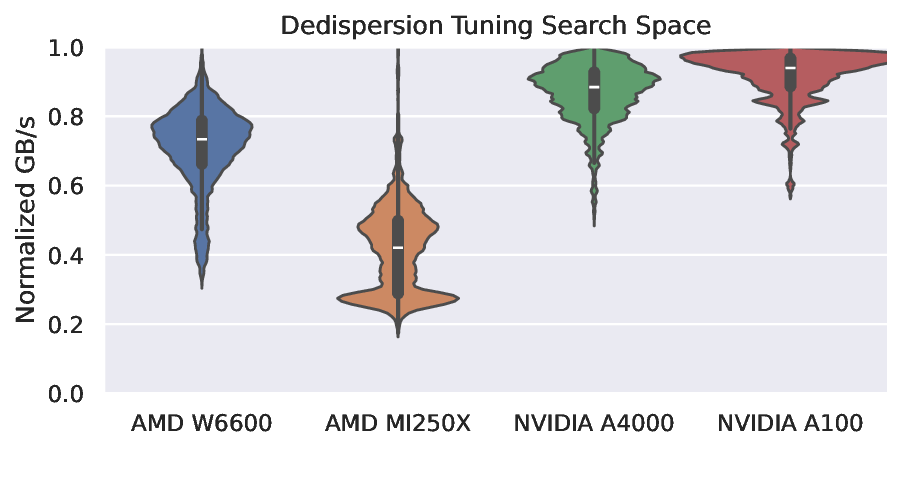}
        \vspace{-0.7cm}
        \caption{Dedispersion tuning search space.}
        \label{fig:dedisp_SP_violins}
    \end{minipage}%
    \begin{minipage}[t]{0.499\textwidth}
        \includegraphics[width=\linewidth]{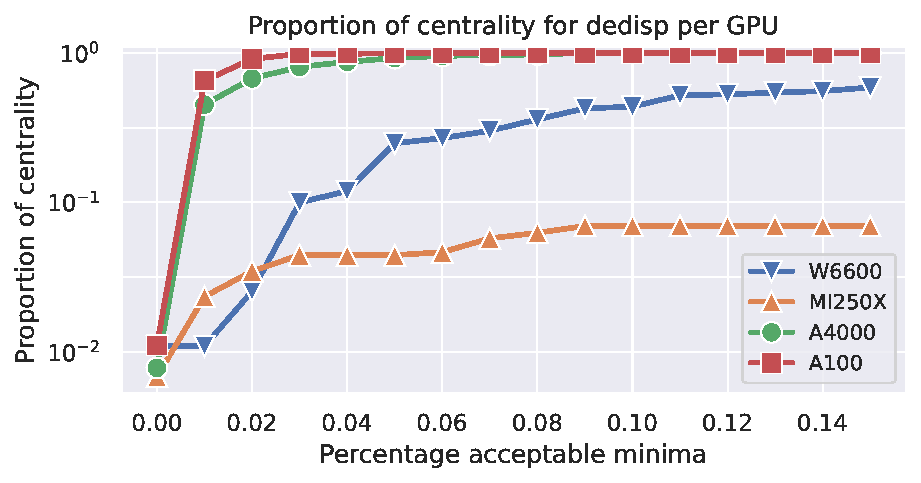}
        \vspace{-0.7cm}
        \caption{Dedispersion proportion of centrality.}
        \label{fig:prop_centr_dedisp}
    \end{minipage}
\vspace{-0.2cm}
    \scriptsize
\captionof{table}{Top configurations for Dedispersion. Parameters match Table~\ref{tab:parameters_conv_hot_dedisp}. Performance in GB/s.}\label{dedisp_params}
\centering
\begin{minipage}{0.249\columnwidth}\scriptsize\centering
W6600 \\
\begin{tabular}{rrrrrr|r}
\toprule
\multicolumn{6}{c|}{Parameters} & Perf. \\
\midrule
32 & 32 & 1 & 1 & 0 & 0 & 582.19 \\ 
2 & 96 & 1 & 1 & 0 & 0 & 575.18 \\ 
16 & 64 & 1 & 1 & 0 & 0 & 573.26 \\ 
2 & 128 & 1 & 8 & 0 & 1 & 568.87 \\ 
4 & 112 & 1 & 1 & 0 & 0 & 568.40 \\ 
\midrule
\end{tabular} \\
\end{minipage}%
\begin{minipage}{0.249\columnwidth}\scriptsize\centering
MI250X \\
\begin{tabular}{rrrrrr|r}
\toprule
\multicolumn{6}{c|}{Parameters} & Perf. \\
\midrule
8 & 32 & 1 & 1 & 0 & 0 & 1586.43 \\ 
8 & 64 & 1 & 1 & 0 & 0 & 1584.01 \\ 
4 & 64 & 1 & 1 & 0 & 0 & 1579.71 \\ 
16 & 32 & 1 & 1 & 0 & 0 & 1576.24 \\ 
4 & 32 & 1 & 1 & 0 & 0 & 1576.03 \\ 
\midrule
\end{tabular} \\
\end{minipage}%
\begin{minipage}{0.249\columnwidth}\scriptsize\centering
A4000 \\
\begin{tabular}{rrrrrr|r}
\toprule
\multicolumn{6}{c|}{Parameters} & Perf. \\
\midrule
8 & 96 & 1 & 6 & 0 & 1 & 532.46 \\ 
8 & 96 & 1 & 4 & 0 & 1 & 532.32 \\ 
16 & 48 & 1 & 5 & 0 & 1 & 532.25 \\ 
8 & 64 & 1 & 5 & 0 & 1 & 532.00 \\ 
8 & 64 & 1 & 7 & 0 & 1 & 531.99 \\ 
\midrule
\end{tabular} \\
\end{minipage}%
\begin{minipage}{0.249\columnwidth}\scriptsize\centering
A100 \\
\begin{tabular}{rrrrrr|r}
\toprule
\multicolumn{6}{c|}{Parameters} & Perf. \\
\midrule
4 & 64 & 1 & 3 & 0 & 1 & 1154.54 \\ 
8 & 96 & 1 & 7 & 0 & 1 & 1153.83 \\ 
8 & 96 & 3 & 7 & 1 & 1 & 1153.06 \\ 
16 & 48 & 1 & 7 & 0 & 1 & 1151.78 \\ 
4 & 64 & 1 & 4 & 0 & 1 & 1151.38 \\ 
\midrule
\end{tabular} \\
\end{minipage}%
\end{figure}

Now we shall look at the Dedispersion kernel. 
In Figure~\ref{fig:dedisp_SP_violins}, we can see a clear distinction in the distribution of the MI250X compared with the other GPUs, where
the optimum is clearly an outlier in terms of performance.
In particular, looking at the median values, the A100 and A4000 achieve respectively the 94\% and 88\% of the optimum, making these devices not difficult to tune manually.

In terms of absolute performance, shown in Table~\ref{tab:result_stats}, we can see that the MI250X achieves the highest overall performance, and over 96\% of its peak bandwidth, and while it is more difficult to tune than the others, the impact is also higher. The proportion of centrality (Figure~\ref{fig:prop_centr_dedisp}) shows that the MI250X remains difficult even if we include more configurations in the acceptable range.
In contrast, the A100 achieves only 74\% of its peak, but the majority of configurations come close to the optimal performance on A100.

Table~\ref{dedisp_params} shows the top configurations on each device for the Dedispersion kernel. One thing that stands out is that all GPUs have a strong preference for large thread blocks, something that we could not have found using OpenCL instead of HIP for the AMD GPUs.
More importantly, all GPUs prefer to do more work in the y-dimension, either per block or per thread, which is the one dimension where data reuse can be exploited.
In particular, the W6600 benefits from its large L3 cache (32MB), achieving up to 582 GB/s, which is more than double of its theoretical peak memory bandwidth.

\subsection{GEMM (General Matrix Multiplication)}

\begin{figure}[tb]
\vspace{-0.2cm}
    \centering
    \begin{minipage}[t]{0.499\textwidth}
        \includegraphics[width=\linewidth]{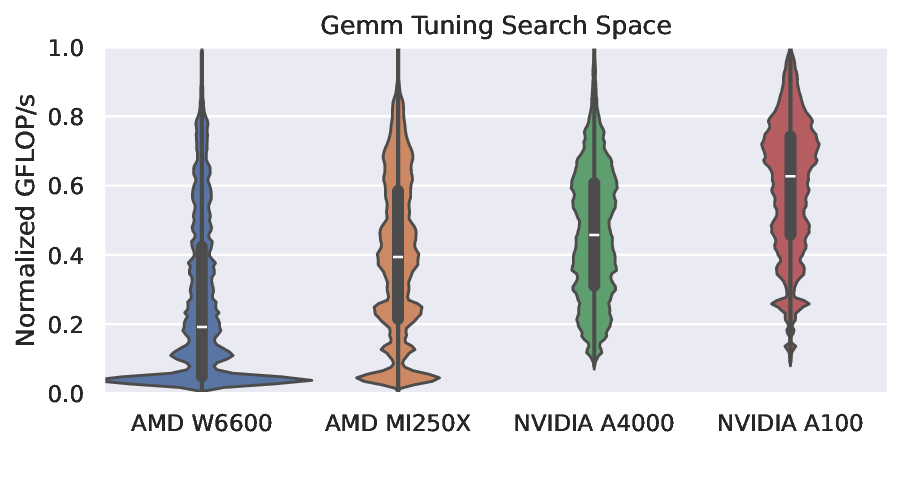}
        \vspace{-0.7cm}
        \caption{GEMM tuning search space.}
        \label{fig:gemm_SP_violins}
    \end{minipage}%
    \begin{minipage}[t]{0.499\textwidth}
        \includegraphics[width=\linewidth]{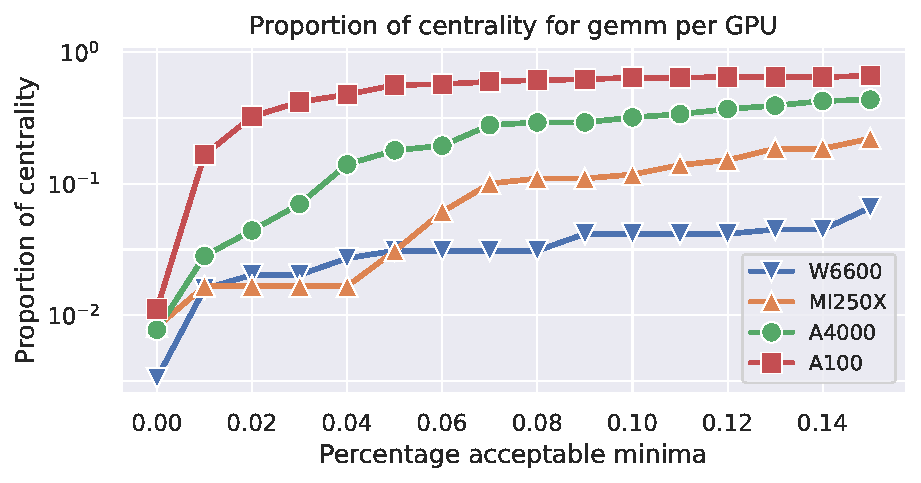}
        \vspace{-0.7cm}
        \caption{GEMM proportion of centrality.}
        \label{fig:prop_centr_gemm}
    \end{minipage}
\vspace{-0.2cm}
    \setlength{\tabcolsep}{.5\tabcolsep}
    \scriptsize
\captionof{table}{Top configurations for GEMM. Parameters match Table~\ref{tab:parameters_gemm}. Performance in GFLOPS/s.}\label{gemm_params}
\centering
\begin{minipage}{0.499\columnwidth}\scriptsize\centering
W6600 \\
\begin{tabular}{rrrrrrrrrrrrr|r}
\toprule
\multicolumn{13}{c|}{Parameters} & Perf. \\
\midrule
128 & 128 & 32 & 32 & 16 & 32 & 16 & 1 & 2 & 1 & 1 & 1 & 1 & 6010.47 \\ 
128 & 128 & 32 & 32 & 16 & 16 & 16 & 2 & 2 & 1 & 1 & 1 & 1 & 6010.17 \\ 
128 & 128 & 32 & 32 & 16 & 32 & 32 & 2 & 2 & 1 & 1 & 1 & 1 & 5992.99 \\ 
128 & 128 & 32 & 32 & 16 & 16 & 32 & 2 & 2 & 1 & 1 & 1 & 1 & 5985.31 \\ 
128 & 128 & 32 & 16 & 32 & 32 & 16 & 2 & 2 & 1 & 1 & 1 & 1 & 5982.45 \\ 
\midrule
\end{tabular} \\
\end{minipage}%
\begin{minipage}{0.499\columnwidth}\scriptsize\centering
MI250X \\
\begin{tabular}{rrrrrrrrrrrrr|r}
\toprule
\multicolumn{13}{c|}{Parameters} & Perf. \\
\midrule
128 & 128 & 16 & 16 & 16 & 32 & 32 & 4 & 2 & 1 & 1 & 1 & 1 & 19806.86 \\ 
128 & 128 & 16 & 16 & 16 & 32 & 32 & 4 & 4 & 1 & 1 & 1 & 1 & 19718.46 \\ 
128 & 128 & 16 & 16 & 16 & 32 & 32 & 4 & 4 & 1 & 0 & 1 & 1 & 19686.09 \\ 
128 & 128 & 16 & 16 & 16 & 16 & 16 & 2 & 2 & 1 & 1 & 1 & 1 & 19651.96 \\ 
128 & 128 & 16 & 16 & 16 & 16 & 16 & 4 & 2 & 1 & 1 & 1 & 1 & 19569.83 \\ 
\midrule
\end{tabular} \\
\end{minipage}
\begin{minipage}{0.499\columnwidth}\scriptsize\centering
A4000 \\
\begin{tabular}{rrrrrrrrrrrrr|r}
\toprule
\multicolumn{13}{c|}{Parameters} & Perf. \\
\midrule
128 & 128 & 16 & 16 & 8 & 8 & 8 & 4 & 4 & 1 & 1 & 1 & 1 & 10502.17 \\ 
128 & 128 & 16 & 8 & 16 & 16 & 16 & 4 & 4 & 1 & 1 & 1 & 1 & 10489.60 \\ 
128 & 128 & 16 & 16 & 8 & 16 & 16 & 4 & 4 & 1 & 1 & 1 & 1 & 10479.14 \\ 
128 & 128 & 16 & 16 & 8 & 16 & 8 & 4 & 4 & 1 & 1 & 1 & 1 & 10469.28 \\ 
128 & 128 & 16 & 8 & 16 & 16 & 8 & 4 & 4 & 1 & 1 & 1 & 1 & 10418.63 \\ 
\midrule
\end{tabular} \\
\end{minipage}%
\begin{minipage}{0.499\columnwidth}\scriptsize\centering
A100 \\
\begin{tabular}{rrrrrrrrrrrrr|r}
\toprule
\multicolumn{13}{c|}{Parameters} & Perf. \\
\midrule
128 & 64 & 16 & 16 & 8 & 32 & 8 & 2 & 4 & 1 & 1 & 1 & 1 & 17145.04 \\ 
64 & 128 & 16 & 8 & 16 & 16 & 16 & 4 & 4 & 1 & 1 & 1 & 1 & 17138.06 \\ 
128 & 64 & 16 & 16 & 8 & 8 & 8 & 4 & 4 & 1 & 1 & 1 & 1 & 17135.74 \\ 
128 & 64 & 16 & 16 & 8 & 16 & 8 & 4 & 4 & 1 & 1 & 1 & 1 & 17123.11 \\ 
64 & 128 & 16 & 8 & 16 & 16 & 8 & 4 & 4 & 1 & 1 & 1 & 1 & 17116.28 \\ 
\midrule
\end{tabular} \\
\end{minipage}%
\end{figure}

Finally, we study the GEMM kernel.
In Figure~\ref{fig:gemm_SP_violins} we notice that the shape of the violin plots for the W6600 and the MI250X are quite similar, although the median performance of the W6600 is barely 20\% of the optimum. The outlier for GEMM is the A100, for which the distribution is more top heavy with half of the configurations within 60\% of the optimum.
However, Table~\ref{tab:result_stats} shows that the speedup over the median is still 1.6x even for the A100. The GEMM kernel on the A100 achieves 88\% of the theoretical peak performance of the GPU. 

Looking at the proportion of centrality in Figure~\ref{fig:prop_centr_gemm} we see that, while the optimal configurations are outliers on all GPUs, including more configurations in the acceptable range makes tuning easier for all devices. The Nvidia GPUs do become easier to tune, compared to AMD, even after a modest increase of the optimality criterion.

Table~\ref{gemm_params} shows again that no single configuration appears in the top 5 for more than one GPU. At the same time, there is a lot of similarity between the top configurations on all four GPUs. For example, all GPUs prefer to store both matrix A and B in shared memory and use a 16 as the loop blocking value for the K loop (KWG, 3rd column in Table~\ref{gemm_params}). 
The thread block dimensions (MDIMC \& NDIMC, 4th and 5th columns) shows that AMD GPUs overall prefer larger thread blocks than the A4000 and the A100. The two server grade GPUs strongly prefer to assign 8 by 8 blocks of work to each thread ($\frac{\text{MWG}}{\text{MDIMC}}$ in x and $\frac{\text{NWG}}{\text{NDIMC}}$ in y dimension), while the top configurations for the A4000 uses 16 in either x or y, and the W6600 uses 4 in the x or y dimension.
We see here the effects of the small cache size of A4000, that prefers to rely on data reuse in registers, compared with the large L3 cache of the W6600, that instead prefers relying more heavily on the hardware managed cache.

\subsection{Performance Portability}

Next, we consider the performance portability of our benchmarks.
Given that the performance portability score \mbox{\PP{}} is computed over a specific set of devices $H$, we can consider different aspects of performance portability by using different subsets of devices for $H$.
For instance, by identifying the configuration with the optimal $\PP{}$ score for $H=\{\text{W6600}, \text{MI250X}\}$ we can determine the most portable configuration across the two AMD devices. 
In this work, we consider the following seven options for $H$:

\begin{itemize}
\item Each of the four GPUs individually.
\item The two AMD devices together: W6600 and MI250X.
\item The two Nvidia devices together: A4000 and A100.
\item All four devices together.
\end{itemize}

\begin{figure*}[tb]
  \centering
  \includegraphics[width=\textwidth]{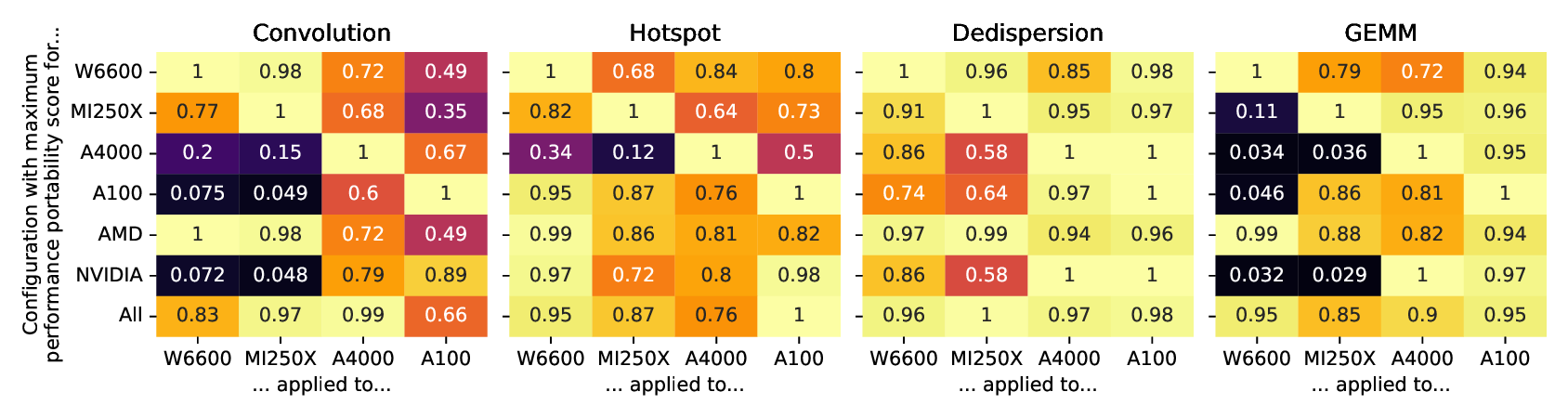}
  \vspace{-0.8cm}
  \caption{Performance portability results. Each row considers a different subset $H$ and shows the results for the configuration $x$ with a maximum $\PP{}(x, p, H)$ score as defined by Eq.~\ref{eq:performance_portability}. Values shown are the application efficiencies $e_i(x, p)$  of $x$ as defined in Eq.~\ref{eq:application_efficiency} for the different devices.\vspace{-.5cm}}
  \label{fig:portability}
\end{figure*}

For each combination of subset $H$ and kernel, we calculated the performance portability $\PP$ across all configurations and selected the one with the highest score.
Figure~\ref{fig:portability} shows the results for each of the three kernels. 
From these results, we can make the following observations.

For the dedispersion and GEMM kernels, we observe that a highly portable configuration exists that achieves an application efficiency of at least 85\% across all devices (bottom row). 
However, for the convolution and the hotspot kernel, we do not find a configuration that qualifies as performance-portable, as each configuration results in a performance loss of at least 15\% on one or more devices.

Another observation is that, in general, configurations performing well on Nvidia tend to not translate to good performance on AMD.
This is especially evident when looking at GEMM and convolution, where configurations exists that obtain more than 80\% of the performance on both Nvidia devices (sixth row), but achieve abysmal performance of less than 10\% on AMD.
Similar patterns can be observed for the other two kernels, albeit with less pronounced differences.
Figure~\ref{fig:portability_avg} shows the average results, revealing that the configuration most portable across Nvidia gives 93\% of the performance on Nvidia and only 41\% on AMD.
These findings underscore the necessity of re-tuning applications previously optimized for Nvidia GPUs when porting to AMD.

However, the converse is not true, and configurations that perform well on AMD typically also perform well on Nvidia.
For example, for GEMM, the configuration that exhibits the highest portability across AMD (fifth row) also delivers 97\% of the performance on the A4000 and 96\% on the A100.
On average, when considering the most portable configurations for AMD across the four kernels, we find AMD gives 97\% of the optimal performance and Nvidia achieves 81\%.

Another observation is that the convolution kernel presents an especially difficult target to tune for, since configurations that perform well on each GPU individually (top four rows), perform poorly on the other devices.
Especially the optimal configurations for the A100, delivers poor performance on AMD.

\section{Discussion}\label{sec:discuss}

In this section, we look at the results of all experiments presented in Section~\ref{sec:evaluation} and provide some highlights on tuning impact, difficulty, and performance portability for all applications and GPUs.

We defined the tuning impact as the performance improvement of the optimum over the median of the tuning space. 
There are clear differences between the impact on performance of auto-tuning on AMD and Nvidia GPUs: the average performance improvement, over all applications, for AMD is 10 times, while for Nvidia it is only 2x.
Our results show that auto-tuning is crucial to achieving high performance for all applications and GPUs in our experiments, but the performance impact is much larger for AMD GPUs than for Nvidia GPUs.

Auto-tuning is not only more important in terms of achieved performance on AMD compared to Nvidia, it is also more difficult. We observe that for all applications the optimum is more of an outlier for AMD than it is for Nvidia.
This does not mean that tuning these applications on the A4000 or A100 is particularly easy, but rather that tuning for the W6600 or the MI250X is, on average, more difficult.

\begin{figure}[t!]
    \centering
    \begin{minipage}[t]{0.49\textwidth}
    \includegraphics[width=\linewidth]{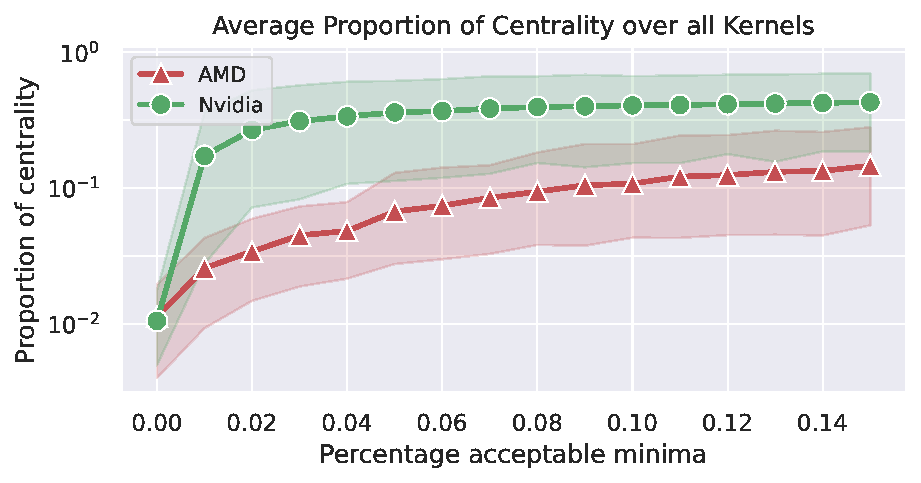}
    \caption{Average proportion of centrality.}
    \label{fig:prop_centr_avg}
    \end{minipage}~%
    \begin{minipage}[t]{0.49\textwidth}
    \centering
    \includegraphics[width=.8\linewidth]{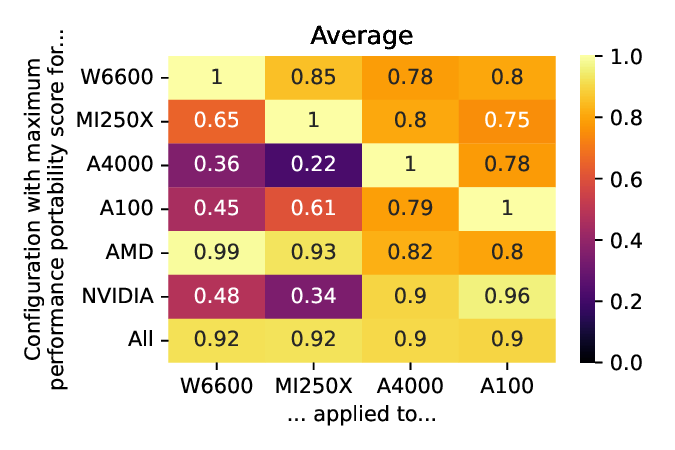}
    \caption{Data from Fig.~\ref{fig:portability} averaged over kernels.}
    \label{fig:portability_avg}
    \end{minipage}
\end{figure}

In Figure~\ref{fig:prop_centr_avg}, we see the averaged proportion of centrality for all the applications, showing that while the global optimum is difficult to find for both vendors, if we relax the constraint on optimality the Nvidia GPUs become easier to tune than the AMD GPUs.
We can conclude that, for our benchmarks, tuning HIP kernels is overall more difficult for AMD than for Nvidia.

By using the performance portability metric, we assessed how well a kernel tuned for one specific GPU performs on the other devices. 
A final observation from Figure~\ref{fig:portability_avg} is that configurations that perform well on the A4000 often fall short on AMD devices. 
On average, the configuration that achieves optimal performance on the A4000, only attains an average performance of ${\sim}22\%$ on the MI250X and ${\sim}36\%$ on the W6600.

\section{Conclusions}\label{sec:conclusion}

In this paper, we compared the auto-tuning effectiveness between AMD and Nvidia GPUs. We integrated support for HIP into Kernel Tuner, now available in the production-ready version 1.0 of the tool, enabling us to auto-tune GPU kernels on both AMD and Nvidia devices. We have compared the impact, tuning difficulty, and performance portability on AMD and Nvidia using four different kernels: 2D convolution, hotspot, dedispersion, and GEMM.

For all four kernels, we see larger differences between the global optimum and the average performance within the search spaces on AMD, compared to Nvidia. This shows that auto-tuning is crucial for achieving high performance on AMD, while manual or no optimization may still yield relatively good performance on Nvidia hardware. Overall, the impact on performance of tuning the same HIP code on AMD GPUs is much larger (10x vs 2x) compared to Nvidia GPUs.

Our evaluation also shows that it is easier for an optimization algorithm to find near-optimal implementations on Nvidia, compared to AMD. Generally, AMD-tuned kernels perform well on Nvidia, but the reverse is not consistently true. 
Thus, while HIP enables \emph{code} portability, it does not guarantee \emph{performance} portability. 
Given that many current GPU applications are written in CUDA and optimized for Nvidia, re-tuning is crucial when migrating to HIP for AMD execution.
Fortunately, the extensions to Kernel Tuner presented in this paper make it possible to tune GPU kernels using HIP on AMD.

This study opens up several avenues for future research. Future work could include a broader array of computational kernels and a broader range of devices from both vendors to fully assess the generalizability of the findings. Also, the disparity in performance portability between Nvidia and AMD GPUs when using HIP suggests a need for deeper investigation into the underlying reasons for these differences. This could involve analyzing the architectural differences between the GPUs of both vendors and how they interact with the HIP programming language. Finally, our extensions to Kernel Tuner bring us one step closer to investigating the effectiveness of auto-tuning for optimizing the energy efficiency of applications on AMD GPUs.

\section*{Acknowledgment and Artifact Availability}

The CORTEX project has received funding from the Dutch Research Council (NWO) in the framework of the NWA-ORC Call (file number NWA.1160.18.316). Funded by the European Union. The ESiWACE3 project has received funding from the European High Performance Computing Joint Undertaking (JU) under grant agreement No 101093054.
The code is available in the repository~\cite{lurati_2024_11617999}.

\bibliographystyle{splncs04}
\bibliography{paper}

\end{document}